\documentstyle[aps,prl]{revtex}

\begin{document}
\title{
Receiver-Operating-Characteristic  Analysis Reveals Superiority of Scale-Dependent Wavelet and
Spectral Measures for Assessing Cardiac Dysfunction
}
\author{
Stefan Thurner,$^1$ 
\footnote{ now at Institut f\"ur Kernphysik, TU-Wien, 
           Wiedner Hauptstrasse 8-10,
           A-1040 Vienna, Austria} 
Markus C. Feurstein,$^1$
\footnote{ now at Dept. of Industrial Information Processing,
           WU-Wien, Pappenheimgasse 35/3/5,
           A-1200 Vienna, Austria} 
Steven B. Lowen,$^1$
and Malvin C.~Teich$^{1,2}$ \\
$^1$ {\it Department of Electrical and Computer Engineering,} \\
{\it Boston University, } {\it  Boston, Massachusetts 02215, USA} \\
$^2$ {\it Departments of Biomedical Engineering and Physics,}\\
{\it Boston University, } {\it  Boston, Massachusetts 02215, USA} \\
}

\maketitle

\begin{abstract}

\noindent

Receiver-operating-characteristic (ROC) analysis was used to assess
the suitability of various heart rate variability (HRV) measures for
correctly classifying electrocardiogram records of varying lengths as
normal or revealing the presence of heart failure. 
Scale-dependent HRV measures were found to be substantially superior
to scale-independent measures (scaling exponents) for discriminating
the two classes of data over a broad range of record lengths.
The wavelet-coefficient standard deviation at a scale near 32
heartbeat intervals, and its spectral counterpart near 1/32 cycles
per interval, provide reliable results using 
record lengths just minutes long.
A jittered integrate-and-fire model built around a fractal
Gaussian-noise kernel provides a realistic, though not perfect,
simulation of heartbeat sequences.
\\
PACS number(s) 87.10.+e, 87.80.+s, 87.90.+y
\end{abstract}

\vspace{0.5cm}

Though the notion of using heart rate variability (HRV) analysis to
assess the condition of the cardiovascular system stretches back some
40 years, its use as a noninvasive clinical tool has
only recently come to the fore \cite{MALI96}. A whole host of
measures, both scale-dependent and scale-independent, have been added
to the HRV armamentarium over the years.

One of the more venerable among the many scale-dependent measures in
the literature is the interbeat-interval (R-R) standard deviation
$\sigma_{\rm int}$ \cite{WOLF78}. 
The canonical example of a scale-independent measure is the scaling
exponent $\alpha_S$ of the interbeat-interval power spectrum,
associated with the decreasing power-law form of the spectrum at
sufficiently low frequencies $f$: $S(f) \propto f^{-{\alpha_S}}$
\cite{MALI96,KOBA82}.
Other scale-independent measures have been developed by us
\cite{TURC93,TURC96,THUR98}, and by others
\cite{PENG93,PENG95,BASS94}. One of the principal goals of this
Letter is to establish the relative merits of these two classes of
measures, scale-dependent and scale-independent, for assessing
cardiac dysfunction.

One factor that can confound the reliability of a measure is the
nonstationarity of the R-R time series.
Multiresolution wavelet analysis provides an ideal means of
decomposing a signal into its components at different scales
\cite{DAUB82,ALDR96,METN97}, and at the same time has the salutary
effect of eliminating nonstationarities \cite{ARNE95,TEIC96}. It is
therefore ideal for examining both
scale-dependent and scale-independent measures; it is in this latter
capacity that it provides an estimate of the wavelet scaling exponent
$\alpha_W$ \cite{THUR98}.

We recently carried out a study \cite{THUR98}
in which wavelets were used to analyze
the R-R interval sequence from
a standard electrocardiogram (ECG) database \cite{data}.
Using the wavelet-coefficient standard deviation
$\sigma_{{\rm wav}}(m)$, where $m=2^r$ is the scale and $r$ is the
scale index, we
discovered a critical scale window near $m=32$ interbeat intervals
over which it was possible to perfectly discriminate
heart-failure patients from normal subjects.
The presence of this scale window was confirmed in an Israeli-Danish
study of diabetic patients who had not yet developed clinical signs
of cardiovascular disease \cite{ASHK98}. These two studies
\cite{THUR98,ASHK98}, in conjunction
with our earlier investigations which revealed a similar critical
scale window in the {\it counting} statistics of the
heartbeat \cite{TURC93,TURC96,TEIC96B}
(as opposed to the time-{\it interval} statistics considered here),
lead to the recognition that scales in the vicinity of $m=32$ enjoy a
special status.
This conclusion has been borne out for a broad range of analyzing
wavelets, from Daubechies 2-tap (Haar) to Daubechies 20-tap
\cite{DAUB82,TEIC99} (higher order analyzing wavelets are suitable
for removing polynomial nonstationarities \cite{ALDR96}).
It is clear that scale-dependent measures
[such as $\sigma_{{\rm wav}}(32)$] substantially outperform
scale-independent ones (such as $\alpha_S$ and $\alpha_W$) in their
ability to discriminate patients with certain cardiac dysfunctions
from normal subjects (see also \cite{TEIC99,TEIC98}).

The reduction in the value of the wavelet-coefficient standard
deviation $\sigma_{{\rm wav}}(32)$ that leads to the scale window
occurs not only for heart-failure patients \cite{THUR98}, but also
for heart-failure patients with atrial fibrillation \cite{TEIC99},
diabetic patients \cite{ASHK98}, heart-transplant patients
\cite{ASHK98,TEIC98}, and in records preceeding sudden cardiac death
\cite{THUR98,TEIC98}.
The depression of $\sigma_{{\rm wav}}(32)$ at these
scales is likely associated with the impairment of autonomic nervous
system function. Baroreflex modulations of the sympathetic or
parasympathetic tone typically lie in the range 0.04--0.09 Hz
(11--25 sec), which corresponds to the time range where
$\sigma_{{\rm wav}}(m)$ is reduced.

The perfect separation achieved in our initial study of 20-h
Holter-monitor recordings endorses the choice of
$\sigma_{{\rm wav}}(32)$ as a useful diagnostic measure. The results
of most studies are seldom so clear-cut, however.
When there is incomplete separation between
two classes of subjects, as observed for other less discriminating
measures using
these identical long data sets \cite{PENG93,PENG95}, or when
our measure is applied to large collections of out-of-sample or
reduced-length
data sets \cite{TEIC98}, an objective means for determining the
relative diagnostic abilities of different measures is required.

{\it ROC~Analysis.}--
\mbox{Receiver-operating-characteristic} (ROC) 
analysis \cite{SWET88} is an
objective and highly effective technique for
assessing the performance of a measure when it is used in binary
hypothesis
testing. This format provides that a data sample be assigned to one
of two hypotheses or classes (e.g., normal or pathologic)
depending on the value of some measured statistic relative to a
threshold value. The
efficacy of a measure is then judged on the basis of its sensitivity
(the proportion of pathologic patients correctly identified) and its
specificity (the proportion of control subjects correctly
identified).
The ROC curve is a graphical presentation of sensitivity versus
$1-$specificity as a threshold parameter is swept (see Fig.~1).

The area under the ROC curve serves as a well-established index of
diagnostic accuracy \cite{SWET88}; a value of 0.5 arises from
assignment to a class by pure chance whereas the maximum value of 1.0
corresponds to perfect assignment (unity sensitivity for all values
of specificity).
ROC analysis can be used to choose the best of a host of different
candidate diagnostic measures by comparing their ROC areas, or to
establish for a single measure the tradeoff between reduced data
length and misidentifications (misses and false positives) by
examining ROC area as a function of record length (see Fig.~2).
A minimum record length can then be specified to achieve acceptable
classification.
Because ROC analysis relies on no implicit assumptions about the
statistical nature of the data set \cite{TEIC99,SWET88}, it is more
reliable and appropriate for analyzing non-Gaussian time series than
are measures of statistical significance such as {\it p}-value and
$d'$ which are expressly designed for signals with Gaussian
statistics \cite{TEIC99}.
Moreover, ROC curves are insensitive to the units employed (e.g.,
spectral magnitude, magnitude squared, or log magnitude); ROC curves
for a measure $M$ are identical to those for any monotonic
transformation thereof such as $M^x$ or $\log(M)$.
In contrast the values of $d'$, and its closely related cousins,
change under such transformations.
Unfortunately, this is not always recognized which leads some authors
to specious conclusions \cite{AMAR98}.

{\it Scale-Dependent vs Scale-Independent Measures.}--
Wavelet analysis provides a ready comparison for scale-dependent and
scale-independent measures since it reveals both.
ROC curves constructed using 75,821 R-R intervals from each of the 24
data sets (12 heart failure, 12 normal) \cite{data}, are presented in
Fig.~1 (left) for the wavelet measure $\sigma_{{\rm wav}}(32)$ (using
the Haar wavelet) as well as for the wavelet
measure $\alpha_W$.
It is clear from Fig.~1 that the area under the $\sigma_{{\rm
wav}}(32)$ ROC curve is unity, indicating perfect discriminability.
This scale-dependent measure clearly outperforms the
scale-independent measure $\alpha_W$ which has significantly smaller
area.
These results are found to be essentially independent of the
analyzing wavelet \cite{THUR98}.

We now use ROC analysis to quantitatively compare the tradeoff
between reduced record length and misidentifications for this
standard set of heart-failure patients using three scale-dependent
and three scale-independent measures.
In the first category are the wavelet-coefficient standard deviation
$\sigma_{{\rm wav}}(32)$, its spectral counterpart $S(1/32)$ \cite
{TEIC99,HENE99}, and the interbeat-interval standard deviation
$\sigma_{\rm int}$.
In the second category, we consider the wavelet scaling exponent
$\alpha_W$, the spectral scaling exponent $\alpha_S$, and a scaling
exponent $\alpha_D$ calculated according to detrended fluctuation
analysis (DFA) \cite{PENG95}.

In Fig.~2 (left) we present ROC area, as a function of R-R interval
record length, using these six measures.
The area under the ROC curves forms the rightmost point in the ROC
area curves.
The file sizes are then divided into smaller segments of length $L$.
The area under the ROC curve is computed for the first such segment
for all 6 measures, and then for the second segment, and so on for
all segments of length $L$.
From the $L_{\rm max} / L$ values of the ROC area, the mean and
standard deviation are computed.
The lengths $L$ employed range from $L=2^6=64$ to $L=2^{16}=65,536$
in powers of two.

The best performance
is achieved by $\sigma_{{\rm wav}}(32)$ and $S(1/32)$, both of
which attain unity area (perfect separation) for sufficiently long
R-R sequences. Even for fewer than 100 heartbeat intervals,
corresponding to {\it just a few minutes of data}, these measures
provide excellent
results (in spite of the fact that both diurnal and nocturnal records
are included).
$\sigma_{\rm int}$ does not perform quite as well. The worst
performance, however, is provided by the three scaling exponents
$\alpha_W$, $\alpha_S$, and $\alpha_D$,
confirming our
previous findings \cite{TURC93,TURC96,THUR98}. Moreover, results
obtained from the
different power-law estimators differ widely \cite{THUR97},
suggesting that there is little merit in the concept
of a single exponent, no less a ``universal" one \cite{AMAR98}, for
characterizing the human heartbeat sequence. In a recent paper
Amaral et al. \cite{AMAR98} conclude exactly the opposite, that the
scaling exponents provide the best performance. This is because they
improperly make use of the Gaussian-based measures $d^2$ and $\eta$,
which are closely related to $d'$, rather than ROC analysis.
These same authors \cite{AMAR98} also purport to glean information
from higher moments of the wavelet coefficients, but such information
is not reliable because estimator variance increases with moment
order.
The results presented here accord with those obtained in a detailed
study of 16 different measures of HRV \cite{TEIC99}.
There are vast differences in the time required to compute these
measures however: for 75,821 interbeat intervals, $\sigma_{{\rm
wav}}(32)$ requires the shortest time (20 msec) whereas  DFA$(32)$
requires the longest time (650,090 msec).

It will be highly useful to evaluate the relative
performance of these measures for other records, both normal and
pathologic.
In particular the correlation of ROC area with severity of cardiac
dysfunction should be examined.

An issue of importance is whether the R-R sequences, and therefore
the ROC curves, arise from deterministic chaos \cite{BASS94}.
We have carried out a phase-space analysis in which {\it differences}
between adjacent R-R intervals are embedded.
This minimizes correlation in the time series which can interfere
with the detection of deterministic dynamics.
The results indicate that the behavior of the underlying R-R
sequences, both normal and pathological, appear to have stochastic
rather than deterministic origins \cite{TEIC99}.

{\it Generating a realistic heartbeat sequence.}--
The generation of a mathematical point process that faithfully
emulates the human heartbeat could be of importance in a number of
venues, including pacemaker excitation.
Integrate-and-fire (IF) models, which are physiologically plausible,
have been developed for use in cardiology.
Berger et al. \cite{BERG86}, for example, constructed an
integrate-and-fire model in which an underlying rate
function was integrated until it reached a fixed threshold,
whereupon a point event was triggered and the integrator reset.
Improved agreement with experiment was obtained by modeling the
stochastic component of the rate function as band-limited fractal
Gaussian noise (FGN), which introduces scaling behavior into the
heart rate, and setting the threshold equal to unity \cite{TURC96}.
This fractal-Gaussian-noise
integrate-and-fire (FGNIF) model has been quite successful 
in fitting a whole host of
interval- and count-based measures of the heartbeat sequence for
both heart-failure patients and normal subjects \cite{TURC96}.
However, it is not able to accommodate the differences observed in
the behavior of $\sigma_{\rm wav}(m)$ for the two classes of data.

To remedy this defect, we have constructed a jittered version of this
model which we dub the fractal-Gaussian-noise jittered
integrate-and-fire (FGNJIF) model \cite{THUR97}.
The occurrence time of each point of the FGNIF is jittered by a
Gaussian distribution of standard deviation $J$.
Increasing the jitter parameter imparts additional randomness
to the R-R time series at small scales,
thereby increasing  $\sigma_{\rm wav}$ at small values of $m$ and,
concomitantly, the power spectral density at large values of the
frequency $f$.
The FGNJIF simulation does a
rather good job of mimicing patient and control data for a number of
key measures used in heart-rate-variability analysis. 
The model is least successful in fitting the interbeat-interval
histogram $p_{\tau}(\tau)$, particularly for heart-failure
patients. This indicates that that a mechanism other than jitter for
increasing  $\sigma_{\rm wav}$ at low scales should be
sought \cite{TEIC99}.

It is of interest to examine the global performance
of the FGNJIF model using the collection of 24 data sets.
To achieve this we carried out FGNJIF simulations using parameters
comparable with the actual data and constructed simulated ROC curves
for the measures $\sigma_{{\rm wav}}(32)$ and $\alpha_W$ as shown
in Fig.~1 (right).
Similar simulations for ROC area versus record length are displayed
in Fig.~2 (right) for the six experimental measures considered.
Overall, the global simulations (right-hand side of Fig.~1 and~2)
follow the trends of the data (left-hand side of Fig.~1 and~2)
reasonably well, with the exception of $\sigma_{\rm int}$.
This failure is linked to the inability of the simulated results to
emulate the observed interbeat-interval histograms.
It will be of interest to consider modifications of the FGNIF model
that might bring the simulated ROC curves into better accord with the
data-based curves.


\begin{figure}
\caption{
ROC curves (sensitivity vs $1-$specificity) for two wavelet-based
measures:
$\sigma_{{\rm wav}}(32)$ which is scale-dependent and
$\alpha_W$ which is scale-independent.
{\it Left}: ROC curves obtained using all 24 data records,
each comprising 75,821 interbeat intervals \protect\cite{data}.
The scale-dependent measure outperforms the
scale-independent one since its ROC area is greater.
{\it Right}: Comparable result
obtained using simulations for the fractal-Gaussian-noise
jittered integrate-and-fire (FGNJIF)
model.
}
\end{figure}

\begin{figure}
\caption{Diagnostic accuracy (area under ROC curve) {\it vs} data
length (number of R-R intervals) for three scale-dependent and three
scale-independent measures (mean $\pm$ one standard deviation).
An area of unity corresponds to the correct assignment of each
patient to the appropriate class.
{\it Left}: $\sigma_{{\rm wav}}(32)$ and $S(1/32)$ provide excellent
performance, attaining unity area (perfect separation) for 32,768 (or
more) R-R intervals.
These measures continue to perform well even as the number of R-R
intervals decreases below 100, corresponding to record lengths just
minutes long. 
The performance of $\sigma_{\rm int}$ is seen to be slightly
inferior.
In contrast, all three scale-independent measures perform poorly.
{\it Right}: Similar results are obtained using 24 simulations of the
FGNJIF model, with the exception of $\sigma_{\rm int}$ (see text).
}
\end{figure}


\begin{thebibliography}{99}
\bibitem{data}
The R-R recordings used in Ref. \cite{THUR98} were drawn from
the Beth-Israel Hospital (Boston, MA) Heart-Failure
Database comprising 12 records
from normal subjects (age 29--64 years, mean 44 years)
and 15 records from severe congestive heart-failure patients (age
22--71 years, mean 56 years).
The recordings were made with a Holter monitor digitized at a fixed
value of 250 Hz.
Three of the 15 heart-failure patients suffered from atrial
fibrillation.
The analysis reported in this Letter eliminates these three
atrial-fibrillation patients,
and makes use of the first 75,821 sinus-rhythm interbeat intervals
from the remaining 12
normal and 12 congestive heart-failure records.
A detailed characterization of each of the records is presented in
Table~1 of Ref.~\cite{TURC96}.
The conclusions reported in this Letter remain intact even when the
three atrial-fibrillation patients are included, as shown elsewhere
\cite{TEIC99}.
\end{thebibliography}
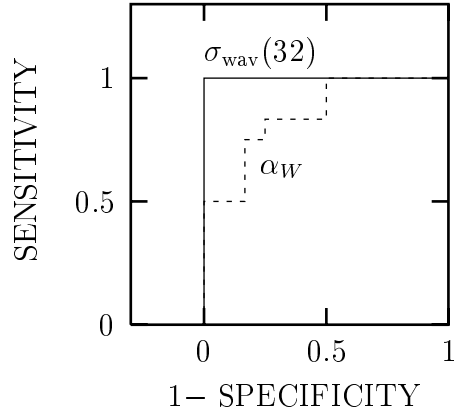
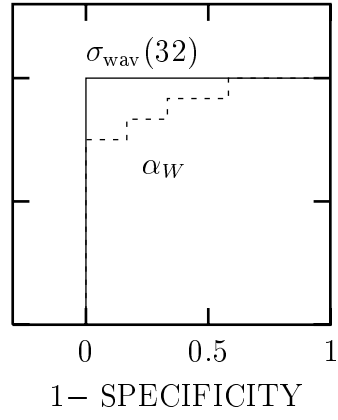
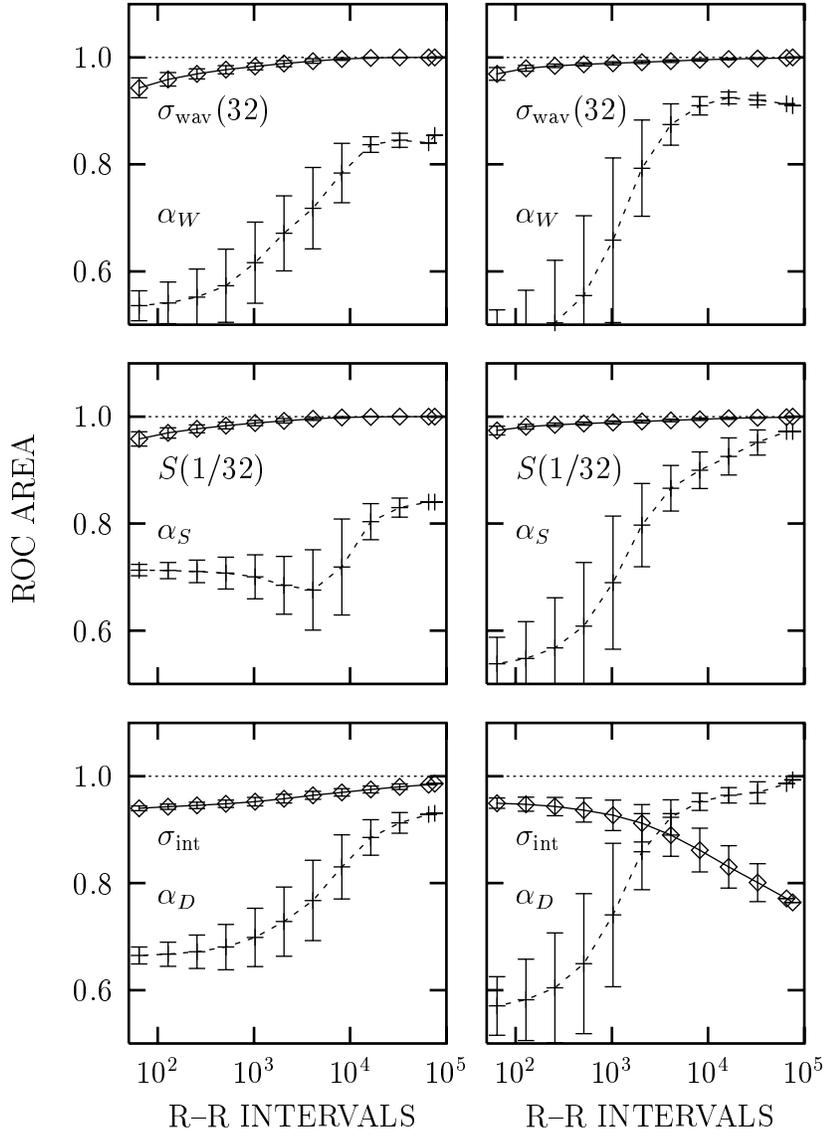
\end{document}